# Gadolinium halide monolayers: a fertile family of two-dimensional 4*f* magnets

Haipeng You,[1] Ning Ding,[1] Jun Chen,[1] Xiaoyan Yao,[1, *] and Shuai Dong[1, *]

[1]*School of Physics, Southeast University, Nanjing 211189, China*

**ABSTRACT:** Two-dimensional (2D) magnets have great potentials for applications in next-generation information devices. Since the recent experimental discovery of intrinsic 2D magnetism in monolayer $CrI_3$ and few-layer $Cr_2Ge_2Te_6$, intensive studies have been stimulated in pursuing more 2D magnets and revealing their intriguing physical properties. In comparison to the magnetism based on 3*d* electrons, 4*f* electrons can provide larger magnetic moments and stronger spin-orbit coupling, but have been much less studied in the 2D forms. Only in very recent years, some exciting results have been obtained in this area. In this mini-review, we will introduce some recent progress in 2D Gd halides from a theoretical aspect. It is noteworthy that 4*f* and 5*d* orbitals of Gd both play key roles in these materials. For Gd$X_2$ (*X*=I, Br, Cl and F) monolayers and related Janus monolayers, robust ferromagnetism with large exchanges comes from the $4f^7 + 5d^1$ hybridization of $Gd^{2+}$. The spatially expanded 5*d* electrons act as a bridge to couple localized 4*f* spins. For Gd$X_3$ monolayers, the intercalation of metal atoms can dope electrons into Gd's 5*d* orbitals, which leads to numerous intriguing physical properties, such as ferroelasticity, ferromagnetism, and anisotropic conductance. In brief, Gd halides establish an effective strategy to take advantage of *f*-electron magnetism in 2D materials.

## 1. INTRODUCTION

In 2004, as a true two-dimensional (2D) material, single-layer graphene was successfully isolated from graphite,[1] thus opening a new door to pursue high-performance 2D materials for diversified device applications.[2-13] During the past decade, the scope of 2D materials has been expanded rapidly, covering numerous famous materials including black phosphorus,[14,15] hexagonal boron nitride (*h*-BN),[16] metal oxides,[17] transition metal dichalcogenides (TMDs),[18-27] and so on. Their excellent physical properties bring more possibilities of further miniaturization and high integratablility for nanoelectronic devices.

In the field of 2D materials, the research on magnetism is always a highlighted topic due to its potential application to spintronics and electronics.[28-35] In 2017, the breakthrough was achieved, that is, the intrinsic ferromagnetism was experimentally discovered in monolayer $CrI_3$ and few-layer $Cr_2Ge_2Te_6$.[36,37] Both of them are ferromagnetic (FM) insulators with a low Curie temperature ($T_C$): namely, $T_C$=45 K for the $CrI_3$ monolayer and $T_C$=28 K for the $Cr_2Ge_2Te_6$ bilayer. Soon afterward, a $Fe_3GeTe_2$ monolayer was reported to be a conductor with intrinsic ferromagnetism, whose $T_C$ value could be modulated by ionic gating.[38] Subsequently, a large number of various 2D magnets, including FM semiconductors (e.g., $MnSe_2$,[27] $VSe_2$[39]), antiferromagnetic (AFM) semiconductors (e.g., $MnPS_3$,[40] $FePS_3$[25]) and FM metals (e.g., $CrB$,[41] $MnX$ ($X$=P, As)[42]), have been synthesized by using mechanical cleavage, molecular beam epitaxy, or chemical vapour deposition.

In terms of the theoretical design of 2D magnets, there are generally two classes of approaches. One approach is to search for 2D intrinsic magnetic materials with magnetic ions, especially van der Waals (vdW) layered magnets.[28,29,43] Then monolayer and few-layer 2D magnets can be obtained by mechanical exfoliation. The other approach is to pursue improper 2D magnetic systems, that is, to introduce spin polarization into nonmagnetic materials. It can obtained via many methods, such as doping with few magnetic atoms, creation of some edge/defects with unpaired electrons, strain engineering, heterojunction construction, and so on.[44-45]

The aforementioned 2D magnets, no matter whether they have been experimentally verified or theoretically predicted, are mostly based on spin moments of $3d$ electrons. Alternatively, here we will provide an overview of recent theoretical progress on another branch, i.e. the 2D Gd halides, and focus on their unique $4f$-electron magnetism and related fascinating physical properties, such as ferrovalley and ferroelasticity. In the following, this review is divided into three parts. The first part will describe the studies on 2D FM Gd$X_2$ ($X$=I, Br, Cl and F).[46-50] The second part will recall Gd dihalide based Janus monolayers, including GdClF, GdBrCl, GdICl and GdIBr.[51,52] The third part will introduce 2D GdI$_3$ and GdCl$_3$ monolayers, as well as their variants with metal atom intercalation.[53,54] It is noteworthy that the $4f$ and $5d$ orbitals of Gd both play crucial roles in these materials. For Gd$X_2$ and related Janus monolayers, the spatially expanded $5d$ electrons act as a bridge to couple localized $4f$ spins, and their $4f^7+5d^1$ hybridization of Gd$^{2+}$ leads to large exchange interactions, as illustrated in Figures 1a-b. For GdI$_3$ and GdCl$_3$ monolayers, the intercalation of Li or Mg atoms can dope electrons into Gd's $5d$ orbitals, which induces a Peierls transition and prominent ferroelasticity, or significant structural reconstruction with ferromagnetism and anisotropic electrical transport.

In comparison with the extensively studied 2D magnets with $3d$ electrons, the $4f$ electrons in 2D materials may provide much stronger spin-orbit coupling (SOC) and much larger local magnetic moment, but have been much less studied to date. Therefore, studies on 2D $4f$-electron magnets are just in the very early stage, with limited literature. Generally speaking, $4f$ orbitals are always too localized spatially, and thus the exchanges between neighboring $4f$ spins are typically weak. It is interesting and important to learn how to use the advantages of $4f$ electrons and circumvent their weakness. In particular, the Gd-based halide monolayers may be the most studied series, with corresponding vdW bulks being given in experiments and theoretical supports. In Comparison with other 4$f$ magnets, some Gd-based magnets can show much higher magnetic ordering temperatures, thanks to the help of its 5$d$ orbitals. Furthermore, many other rare-earth elements, with non-half-filled 4$f$ orbitals, are hard to treat in first-principles calculations. For these reasons, our current mini-review will focus mainly on Gd-based halides, which is a starting point for the studies of 2D 4$f$ magnets. Our mini-review can shed light on the manipulation of the 4$f$-electron magnetism in 2D systems and its related applications.

## 2. GADOLINIUM DIHALIDE: Gd$X_2$ ($X$=I, Br, Cl and F) monolayer

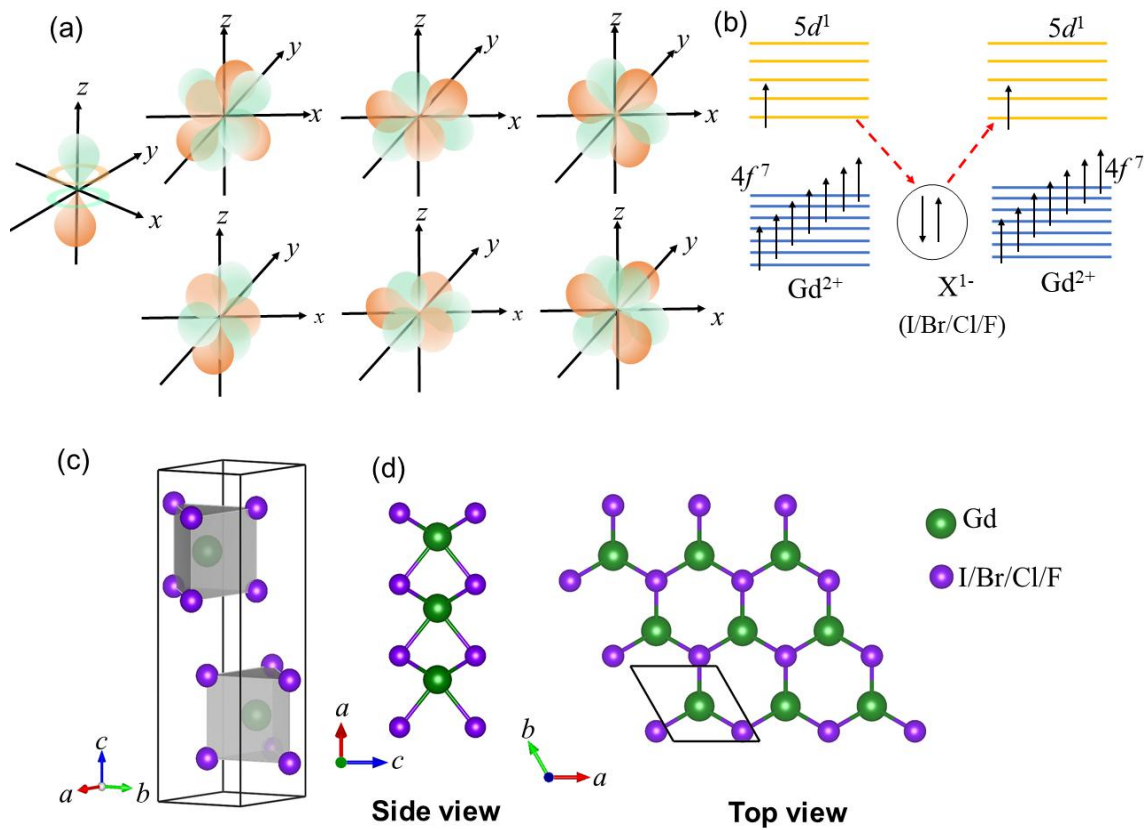

**Figure 1.** (a) Graphic of 4$f$ orbits. (b) Schematic diagram of a double-exchange-like mechanism for Gd$X_2$ ($X$=I, Br, Cl and F) monolayers. (c) Crystal structure of vdW layered bulk GdI$_2$ with the

$P6_3/mmc$ phase. (d) Side and top views of monolayer Gd$X_2$.

Bulk GdI$_2$ with a vdW layered structure was first synthesized by Mee and Corbett *et al*. in 1965,[55] and it possesses a FM order with $T_C$ value up to room temperature (300-340 K).[56-58] As shown in Figure 1c, the GdI$_2$ bulk consists of I-Gd-I sandwich layers with the 2H-MoS$_2$-type structure, and these layers stack along the *c*-axis in an AB sequence. Each Gd ion is caged within a triangular prism formed by six I ions, and in each layer Gd ions constitute a triangular lattice (Figure 1d). Except for GdI$_2$, no experiment has been reported so far regarding other Gd$X_2$ species (*X*=Br, Cl, and F).

In 2020, 2D monolayer GdI$_2$ was predicted to be a FM semiconductor with a $T_C$ value close to ambient temperature.[46] Following this pioneering work, other Gd$X_2$ (*X*=Br, Cl, and F) monolayers with the identical in-plane structure (Figure 1d) were explored by theoretical calculations.[47-49] Their stabilities were confirmed by elastic property calculations and *ab initio* molecular dynamics (AIMD) simulations. The estimated cleavage energy for GdI$_2$ is 0.26 J/m$^2$, which is less than that of graphite, indicating the experimental feasibility of exfoliating a GdI$_2$ monolayer from its layered bulk.[46] The cohesive energies 5.52, 4.42, and 4.01 eV per atom for GdF$_2$, GdCl$_2$, and GdBr$_2$, respectively, are smaller than that of graphene but larger than that of Cu$_2$Ge, implying the great possibility for them to be captured experimentally.[48]

According to the theoretical calculations,[46-50] all Gd$X_2$ monolayers are FM semiconductors with a large magnetic moment 8 $\mu_B$/fu (fu = formula unit), and they all display a typical bipolar magnetic feature. With the GdI$_2$ monolayer as an example, as plotted in Figure 2a, the top valence band and the bottom conduction band possess an inverse spin-polarization orientation, enabling the feasibility to gain half-metallicity and 100% spin-polarized carriers by modulating the position of the Fermi level.[46,50] A calculation of the magnetocrystalline anisotropy energy (MAE) demonstrated that the magnetic anisotropy of GdX$_2$ changes from an out-of-plane preference to an in-plane type for $X$ varying from F to I. The GdI$_2$ monolayer exhibits an easy magnetic plane, while GdF$_2$ and GdCl$_2$ monolayers have an out-of-plane easy axis, as illustrated in Figures 2d-f. By using a Monte Carlo (MC) simulation under the energy mapping scheme, all GdX$_2$ monolayers were predicted to possess a large $T_C$ value close to ambient temperature, namely 241 K or 251 K for GdI$_2$,[46,47] 229 K or 225 K for GdBr$_2$,[47,48] 224 K or 245 K for GdCl$_2$,[47,48] and 300K for GdF$_2$,[48] which are promising for spintronic devices.

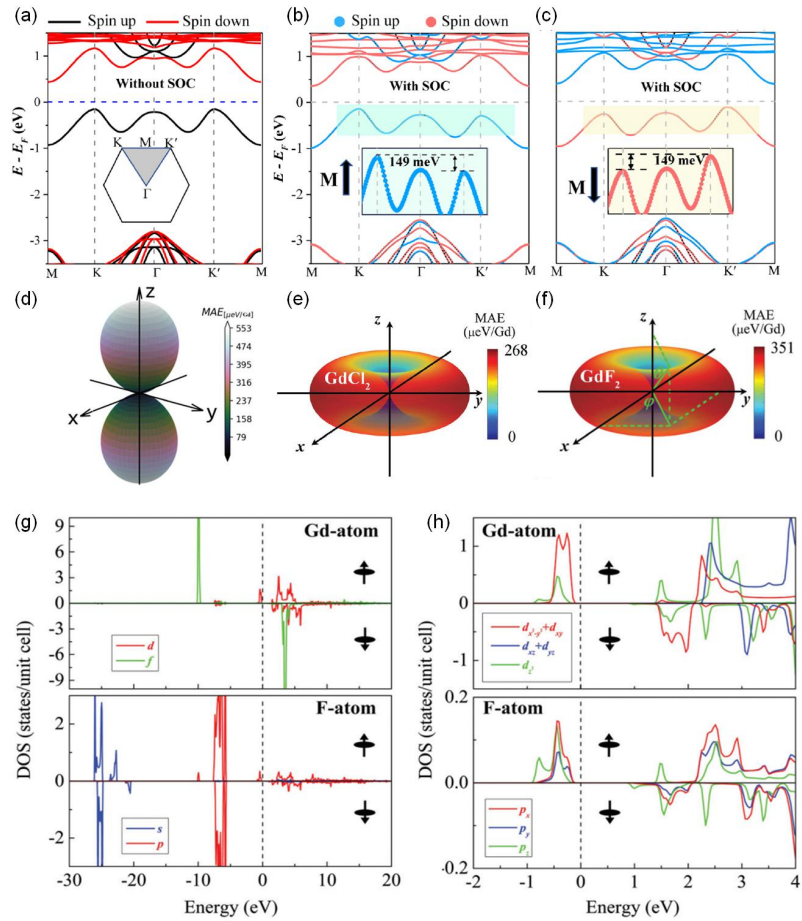

**Figure 2.** Electronic band structure of GdI$_2$ monolayer (a) without SOC; (b, c) with SOC for Gd's magnetic moment along the out-of-plane positive and negative $z$ directions, respectively. (d) Angular dependence of MAE for the GdI$_2$ monolayer with the in-plane MAE set to zero. (e,f)

Angular dependence of MAE for the monolayer $GdCl_2$ and $GdF_2$ respectively with the z-direction MAE set to zero. (g, h) Orbital-resolved density of states for monolayer $GdF_2$ with a FM ground state. The Fermi level is set to zero. Parts (a-c) are reproduced with permission from ref 50. Copyright 2021 American Physical Society. Part (d) is reproduced with permission from ref 46. Copyright 2020 Royal Society of Chemistry. Parts (e-h) are reproduced with permission from ref 48. Copyright 2022 Royal Society of Chemistry.

The intrinsic ferromagnetism with sizable MAE and high $T_C$ is unexpected for the $GdX_2$ monolayers. With the $GdF_2$ monolayer as an example as illustrated in Figure 2g, Gd's $4f$ electrons occupy the deep energy levels far away from the Fermi level. These $4f$ electrons are highly localized due to their narrow and high density distributions. The unexpected ferromagnetism originats from the combined effect of the intra-atomic $Gd_{4f}$-$Gd_{5d}$ coupling and the interatomic $Gd_{5d}$-$Gd_{5d}$ coupling.[46-50] In detail, the Gd ion with a $4f^7 5d^1$ valence electronic configuration contributes a large intrinsic magnetic moment due to the strong Hund interaction between $4f$ and $5d$ electrons. The top of valence band is primarily attributed to Gd's $5d$ orbitals, which appreciably hybridize with $p$ orbitals of $X$ ions, as plotted in Figure 2h. Therefore, the spatially expanded $5d$ electron acts as a bridge to connect localized $4f$ electrons, and the robust ferromagnetism is due to effective $Gd_{4f}$-$Gd_{4f}$ interaction mediated by the $Gd_{5d}$-$X_p$-$Gd_{5d}$ exchanges (Figure 1b). The Gd-$X$-Gd bond angle of close to 90º prefers FM superexchange coupling according to the Goodenough-Kanamori-Anderson (GKA) rule.[59-61] From another perspective, on the basis of a Kramers' mechanism, the partially occupied states also lead to a FM direct-exchange interaction between the nearest-neighboring Gd spins.[60]

In addition to charge and spin, the energy valley degrees of freedom have recently attracted considerable attention,[2,62-64] which can also be used as the basis of information coding. The calculations indicated that all Gd$X_2$ ($X$=I, Br, Cl and F) monolayers are ferrovalley materials that have promise for valleytronic applications.[48-50,64] The spontaneous valley splitting reaches 55 or 47.6 meV for GdF$_2$, 38, 35 or 42.3 meV for GdCl$_2$, and 82 or 79 meV for GdBr$_2$, respectively.[48,49,64] In particular, a giant valley splitting of 149 meV was predicted in GdI$_2$ due to its intrinsic ferromagnetism and large SOC.[50] Furthermore, this valley splitting of GdI$_2$ may be even enhanced to about 189 meV at 10% tensile strain.[50] As exhibited in Figures 2b,c, the valley polarization can be flipped by reversing the spin orientation through a moderate external magnetic field. By applying a suitable external electric field, the anomalous valley Hall effect could be realized.[48] It is worth noting that the energy valley in these Gd$X_2$ monolayers is mainly contributed by the $d$ electrons of Gd, and a small part from the $p$ electrons of $X$.[49] Hence, the electrons on Gd's 5$d$ orbitals and the Gd$_{5d}$-$X_p$-Gd$_{5d}$ FM superexchange play important roles in the valley polarization.

### 3. JANUS MONOLAYER BASED ON GADOLINIUM DIHALIDE

Ever since MoSSe was synthesized successfully,[65,66] 2D Janus-type structures have been widely studied. Recently, a few Janus monolayers based on Gd dihalides were predicted by first-principles calculations, i.e. GdClF, GdBrCl, GdICl and GdIBr monolayers,[51,52] which are represented by Gd$YZ$ in this paper for convenience of description.

The Janus monolayer Gd$YZ$ consists of a $Y$-Gd-$Z$ sandwich layer akin to the aforementioned Gd$X_2$ monolayers, as plotted in Figure 3a. Because of the different atomic sizes and electroegativities of $Y$ and $Z$, the Gd$YZ$ monolayer displays inequivalent Gd-$Y$ and Gd-$Z$ bond lengths, which break the vertical mirror symmetry, and bring about a built-in electric field. The AIMD stimulations and phonon spectra confirmed thermodynamic stability of the Gd$YZ$ monolayers.[51,52] The cleavage energies of GdBrCl, GdICl and GdIBr are 0.203, 0.209 and 0.201 J/m$^2$ respectively,[51] which are smaller than that of graphite, demonstrating the feasibility of exfoliating these Gd*YZ* monolayers from their layered bulks.

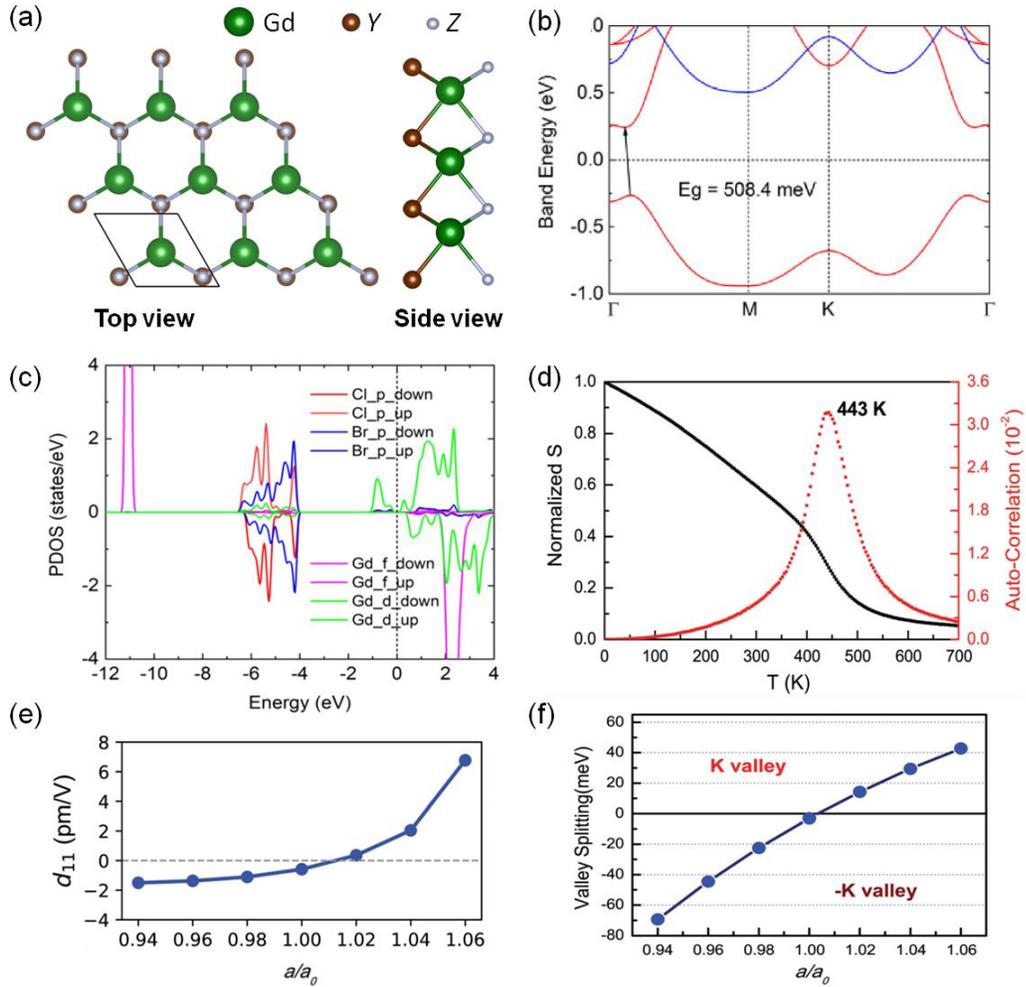

**Figure 3.** (a) Top and side views of the structure for the Janus monolayer Gd*YZ*. (b) Electronic band structure and (c) the spin-polarized atom-orbital-projected density of states for the Janus monolayer GdBrCl without the SOC effect. (d) Normalized magnetic moment (*S*) and autocorrelation of the Janus monolayer GdClF as a function of temperature *T*. (e) The piezoelectric strain coefficient $d_{11}$ and (f) the valley splitting as a function of the biaxial strain $a/a_0$ for the Janus monoalyer GdClF, where *a* and $a_0$ are the strained and unstrained lattice constants, respectively. Parts (b) and (c) are reproduced with permission from ref 51. Copyright 2021 American Institute of Physics.   (d-f) are reproduced with permission from ref 52. Copyright 2022 Royal Society of

Chemistry.

All four of these Gd*YZ* monolayers are FM semiconductors with indirect band gaps. The FM order is stable against biaxial strain in the range of -5% to 5%.[51,52] Except for the GdClF monolayer still retains a bipolar magnetic semiconductor characteristic, the other three Gd*YZ* monolayers show the band gap only being determined by the spin-up channel, as illustrated in Figure 3b with GdBrCl as an example. They inherit the intrinsic ferromagnetism of the same origin as for Gd$X_2$: namely, the magnetic moment with 8 μ$_B$/fu mainly comes from half-filled 4$f$ orbitals and partially occupied 5$d$ orbitals (4$f^7$+5$d^1$). The top of the valence band is primarily attributed to Gd's 5$d$ orbitals, which appreciably hybridize with the $p$ orbitals of *Y* and *Z*, playing a key role in their intrinsic ferromagnetism, as plotted in Figure 3c with GdBrCl as an example. Except for GdClF with a perpendicular magnetic anisotropy, GdIBr, GdICl and GdBrCl monolayers prefer an in-plane magnetic anisotropy. Meanwhile, the strength of the magnetic anisotropy is enhanced by the Janus structure. The MAE values of the GdIBr and GdICl monolayers were predicted to be 708.25 and 613.25 μeV/Gd respectively,[51] which are larger than that of GdI$_2$ (553 μeV/Gd).[46] On the basis of MC simulations, $T_C$ was predicted for these Janus monolayers, namely GdIBr at 167 K, GdICl at 172 K, GdBrCl at 181 K,[51] and a quite high 443 K for GdClF as plotted in Figure 3d,[52] which is desirable for spintronic devices.

In addition, the GdClF monolayer is also a potential ferrovalley material.[52] Although its valley splitting is only -3.1 meV, the valley polarization can be effectively modulated by biaxial strain. The compressive strain can enhance -K valley polarization, while the tensile strain may give rise to K valley polarization. Under a moderate 0.96 compressive strain and 1.04 tensile strain, the corresponding valley splitting energies reach -44.5 and 29.4 meV,[52] respectively (Figure 3f), which can induce an anomalous valley Hall effect tunable by strain. Simultaneously, the strain change from a compressive to a tensile strain can reverse the direction of in-plane piezoelectric polarization (Figure 3e), supplying the probability for a combination of valley properties and piezoelectricity.[52]

## 4. GADOLINIUM TRIHALIDE

In 1960s, bulk GdI$_3$ was experimentally synthesized,[67] which has a vdW layered structure with hexagonal phase (R-3), as displayed in Figure 4a. The I-Gd-I sandwich layers with 1T-BiI$_3$-type structure are stacked along the c axis in an ABC sequence. Each Gd ion is caged within an octahedron formed by eight I ions, while the neighboring octahedrons connect in an edge-sharing manner. In comparison with the cleavage energy of graphite, GdI$_3$ has a much lower value (0.12 J/m$^2$),[53] which suggests that a GdI$_3$ monolayer with an identical in-plane structure (Figure 4b) can be easily exfoliated from its bulk.

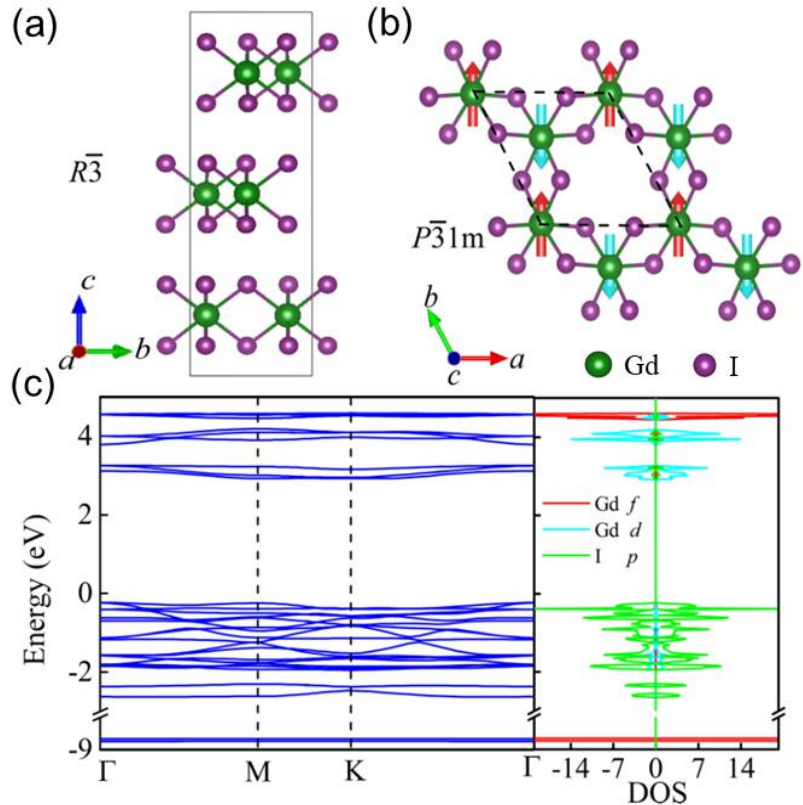

**Figure 4.** (a) Crystal structure of bulk GdI$_3$. (b) Top view of the GdI$_3$ monolayer with a Néel-type-

AFM ground state. (c) Band structure and density of states of the GdI$_3$ monolayer. Reproduced with permission from ref 53. Copyright 2021 American Physical Society.

For the GdI$_3$ monolayer, the Néel-type-AFM state (Figure 4b) was verified to be the ground state,[53] which can be easily understood because the half-filled 4$f$ orbitals prefer an AFM coupling according to the Goodenough-Kanamori rule.[59,60] The calculated Gd magnetic moment is very close to the expected value of 7 $\mu_B$/Gd, which comes from its half-filled 4$f$ orbitals. These 4$f$ orbitals of Gd are so localized that the energy difference between FM and AFM states is very small, inducing a quite low Néel temperature. As displayed in Figure 4c, the Néel-type-AFM GdI$_3$ monolayer is a Mott insulator with fully split narrow 4$f$ bands due to a large Hubbard repulsion. The lowest conducting band is contributed by Gd's 5$d$ orbitals, but the highest valence band is contributed by I's 5$p$ orbitals. In comparison with the 4$f^7$+5$d^1$ hybridization in GdI$_2$,[46] GdI$_3$ only possesses a 4$f^7$ electronic configuration. The lack of 5$d$ electrons cannot supply an effective medium for Gd$_{4f}$-Gd$_{4f}$ coupling, which is the reason for such a weak exchange. The MAE value is only -0.03 meV/Gd, indicating a weak magnetic anisotropy with an out-of-plane easy axis.[53]

For spintronics applications, it is necessary to tune the magnetism of a GdI$_3$ monolayer. Unfortunately, the usual modulation methods of 2D materials, such as biaxial strain, do not work well on the magnetic properties of the GdI$_3$ monolayer due to its rather localized 4$f$ orbitals. To connect these highly localized 4$f$ orbitals, one efficient route is to introduce electrons to Gd's empty 5$d$ orbitals. Different from the close-packed triangular structure in GdI$_2$, the honeycomb configuration in GdI$_3$ provides large interstitial positions of hexa-atomic rings to accommodate other atoms. Thereby, electrons doped into Gd's 5$d$ orbitals can be realized by the intercalation of metal atoms.

When Li atoms are inserted into the interstitial positions of GdI$_3$'s hexatomic rings, as shown in Figure 5a, the chemical formula becomes (GdI$_3$)$_2$Li, which is akin to the intrinsic multiferroic (CrBr$_3$)$_2$Li monolayer.[24] Meanwhile, the valence of Gd become +2.5 and its magnetic moment is ~7.5 μ$_B$/Gd. The MAE value of (GdI$_3$)$_2$Li is estimated to be 0.46 meV/Gd, preferring an in-plane orientation.[53] The enhanced magnetic anisotropy is due to the strong SOC of Gd's partially filled 5$d$ orbitals.

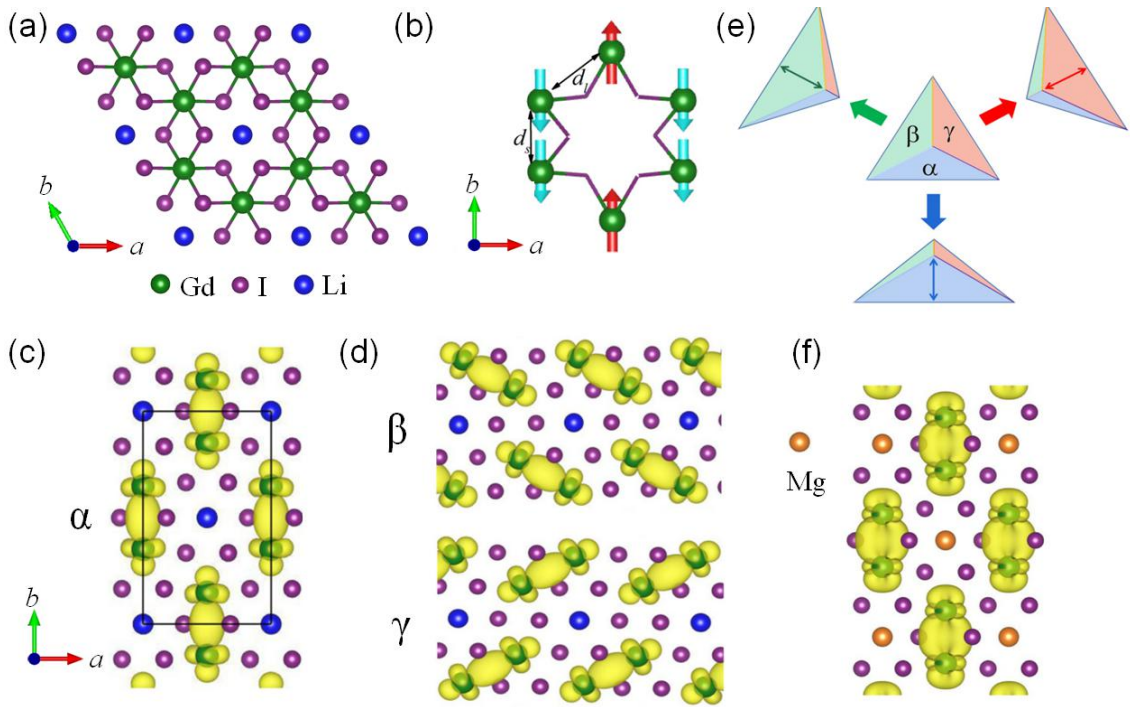

**Figure 5.** (a) Crystal structure of (GdI$_3$)$_2$Li monolayer. (b) Gd framework for the stripy-AFM phase. The structural dimerization can be visualized clearly: the shorter bond between parallel-spin pairs and the longer bond between antiparallel-spin pairs. (c) Schematic of Gd's 5$d$-electron distribution for the (GdI$_3$)$_2$Li monolayer, where the electron spindles are along the $b$ axis in the α domain. (d) Two other ferroelastic domains (β and γ domains) with electron spindles along different directions. (e) Deformation of the macroscopic shape of the (GdI$_3$)$_2$Li monolayer by external forces. (f) Distribution of electrons for the (GdI$_3$)$_2$Mg monolayer. Reproduced with permission from ref 53. Copyright 2021 American Physical Society.

It is interesting that the one electron introduced from a Li atom generates a Periels transition, namely a strong disproportion of the nearest-neighbor Gd-Gd distances induced by dimerization (Figure 5b), which reduces the symmetry from a honeycomb (*P*-31*m*) to a monoclinic phase (*C*2/*m*), and triggers a magnetic transition from Néel-type to Stripy-type AFM state. Due to the strong electron-phonon coupling, a significant ferroelasticity is driven by the Periels transition, making (GdI$_3$)$_2$Li a multiferroic monolayer. The Gd's 5*d* electrons prefer to stay in the middle of the shorter Gd-Gd pair, forming a bond-centered charge ordering such as spindles, as displayed in Figure 5c. This electron spindle leads to a shrunk lattice constants, and thus generates ~4% ferroelastic deformation.[53] Due to the threefold rotational symmetry, the other two similar ferroelastic domains with different orientations of electron spindles are allowed (Figure 5d). The detailed balance among these triple ferroelastic domains can be tuned by external forces along certain directions to reduce mechanical damage, that is, the inner strain can be relaxed by rotating the electron spindles and resizing the corresponding ferroelastic domains, as displayed in Figure 5e. This superelasticity offers a great potential for flexible applications.

After Li intercalation, each Gd gains half of an electron from the Li atom. After Mg intercalation, each Gd gains one electron from the Mg atom. In the (GdI$_3$)$_2$Mg monolayer, as expected, the valence of Gd become +2 and the magnetic moment is ~8$\mu_B$/Gd.[53] However, the ground state still retains a Stripy-AFM phase. Due to more electrons on Gd's 5*d* orbitals, the value of MAE is enhanced to 1.05 meV with an in-plane preference. Although one electron is doped to each Gd, the multiple 5*d* orbitals of Gd separate one electron into a half-occupation of two orbitals, as illustrated in Figure 5f, which also leads to the Periels transition and superior ferroelasticity (~9%).[53]

In addtion to GdI$_3$, the vdW bulk GdCl$_3$ was synthesized experimentally in the 1960s.[68] Since

Cl ions have a stronger electronegativity and smaller ionic radius in comparison to I ions, the Gd-Cl bond is shorter than the Cd-I bond. Hence, the ion cage wrapping each Gd is more compact in GdCl$_3$, forming a rare Cl-hendecahedron, which is different from the common octahedron or triangular prism, as shown in Figure 6a. According to our calculations,[54] the GdCl$_3$ monolayer can be easily exfoliated from its bulk. The GdCl$_3$ monolayer possesses a unique orthorhombic structure with hendecahedral ion cages, which is completely different from both the honeycomb structure of GdI$_3$ and the triangular structure of GdI$_2$, representing a brand-new lattice for a 2D magnet.

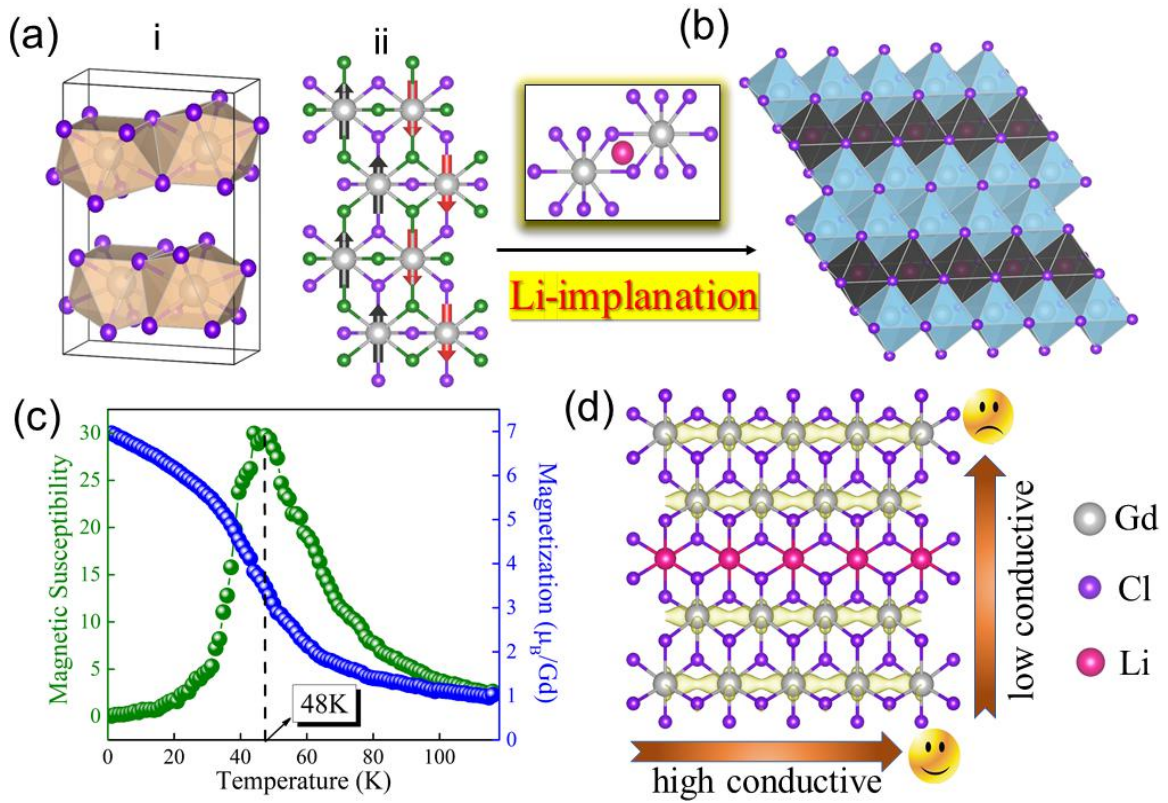

**Figure 6**. (a) (i) Crystal structure of bulk GdCl$_3$ (the *Cmcm* phase) and (ii) a GdCl$_3$ monolayer with a Zigzag-type AFM ground state. (b) Structure of (GdCl$_3$)$_2$Li after reconstruction induced by the Li implanation. (c) Monte Carlo simulated magnetic susceptibility and magnetization. (d) Electronic density distribution of (GdCl$_3$)$_2$Li. Reproduced with permission from ref 54. Copyright 2022 Elsevier Ltd.

GdCl$_3$ monolayer is unique due to not only its structure, but also its magnetism. According to the DFT calculation, it is a Mott insulator with a Zigzag-type AFM ground state (Figure 6a).[54] By insertion of Li into a Gd-Cl$_2$-Gd interstitial position, a significant structural reconstruction occurs, and exotic Cl hendecahedron cages transform to Cl-octahedra (Figure 6b).[54] The symmetry of the resultant (GdCl$_3$)$_2$Li monolayer is reduced to monoclinic (C2/m). Simultaneously, ferromagnetism is achieved with a large magnetization (7.5 μ$_B$/Gd) and a moderate $T_C$ value (~48 K) (Figure 6c).[54] Here, the ferromagnetism is induced by doping electrons in Gd's 5$d$ orbitals, via a double-exchange-like process. Namely, the localized 4$f$ moments act as the spin background while the itinerant 5$d$ electrons act as the medium. Moreover, due to the structural anisotropy, the 5$d$ electrons prefer to stay between short Gd-Gd pairs along the $a$ axis, leading to a large electrical transport anisotropy (Figure 6d), which may be peculiarly useful for nanoelectronics.[54]

## 5. CONCLUSION AND OUTLOOK

In this article, we have reviewed a few representative 2D monolayers based on Gd halides with intrinsic 4$f$ magnetism, including the trigonal FM Gd$X_2$ family, Janus monolayer Gd$YZ$, honeycomb Néel-type-AFM GdI$_3$ and orthorhombic zigzag-type-AFM GdCl$_3$. For Gd$X_2$ and the related Janus monolayer Gd$YZ$, they are all FM semiconductors with high $T_C$ values, which make them promising candidates for spintronic devices. The strong ferromagnetism comes from Gd's localized 4$f$ spins coupled via the spatially expanded 5$d$ electrons. Meanwhile, Gd$X_2$ and GdClF are ferrovalley materials, and their spontaneous valley polarizations are mainly contributed by Gd's 5$d$ electrons. For a GdI$_3$ monolayer without an electron on Gd's 5$d$ orbitals, doping a half or single electron into Gd's 5$d$ orbitals can be achieved by Li or Mg intercalation, which leads to a Peierls transition and prominent ferroelasticity. For a GdCl$_3$ monolayer with an exotic structure, Li implanation also dopes electrons into Gd's 5$d$ orbitals, resulting in a significant structural reconstruction, 4*f* ferromagnetism and distinctive conductance anisotropy.

In compatrison with magnetic materials based on 3*d* electrons, the magnets based on 4*f* electrons may provide unique superiorities, such as large local magnetic moment, excellent magnetic anisotropy, and reliable valley polarization, which will supply more possibilities for their use in multifunctional spintronic and valleytronic devices. For this very young topic of 2D 4*f*-magnets, there is still a long way to go before practical applications can be carried out. Although all the studies mentioned above are only theoretical predictions, they imply an effective strategy to take advantage of the highly localized *f* electrons to improve the magnetism: that is, utilizing the electrons on *d* orbitals as medium. This strategy is not limited to Gd halides, but is also effective for other Gd-based 2D materials.[69]

Experimentally, monolayers of Gd halides can be obtained using two methods. One method is to synthesize their vdW bulks according to well-established chemical synthesis methods and then to exfoliate monolayers by a mechanical method. The other method is to grow monolayers of Gd halides directly by means of molecular beam epitaxy or vapor transport under gas protection.

In addition to Gd halides, there are also some other 4*f* magnets existing experimentally which may also have vdW bulks, such as PrI$_3$, NdI$_3$, TbI$_3$ and so on.[67] However, on the theoretical side, it is usually hard for DFT codes such as VASP to deal with their non-half-filled 4*f* electrons. Thus, up to now the theoretical exploration on these 2D 4*f* magnets has been very limited.[70] Recently, the electronic structure and magnetic property of Eu$X_2$ (*X*=I, Br, C, F) was obtained, since it also has half-filled 4$f^7$ orbitals. In addition to its technical convenience, a Gd-based system is the best choice to pursue high-temperature magnetism, on consideration of its cooperative 5*d* orbitals, which can boost the exchanges between neighboring 4*f* spins. This is the reason the Gd-based halides have been the first attractive species among all 2D 4$f$ magnets. Even so, more 2D 4$f$ magnets with other rare-earth elements are expected to be studied in the future, with the development of both experimental and theoretical techniques.

## ■ AUTHOR INFORMATION

***Corresponding Author**

**Xiaoyan Yao** − *School of Physics, Southeast University, Nanjing 211189, China;*

Email: yaoxiaoyan@seu.edu.cn

***Corresponding Author**

**Shuai Dong** − *School of Physics, Southeast University, Nanjing 211189, China; orcid.org/0000-0002-6910-6319;* Email: sdong@seu.edu.cn

**Authors**

**Haipeng You** − *School of Physics, Southeast University, Nanjing 211189, China*

**Ning Ding** − *School of Physics, Southeast University, Nanjing 211189, China*

**Jun Chen** − *School of Physics, Southeast University, Nanjing 211189, China*

This work was supported by the National Natural Science Foundation of China (Grant No. 11834002). We thank the Tianhe-II of National Supercomputer Center in Guangzhou (NSCC-GZ) and the Big Data Center of Southeast University for providing the facility support on the numerical calculations.